\documentclass[cits]{PoS}

\usepackage[utf8x]{inputenc}
\usepackage{mciteplus}
\usepackage{amsmath}

\newcommand{\rt}{{\mathbf{r}_T}}

\newcommand{\bt}{{\mathbf{b}_T}}

\newcommand{\ud}{\, \mathrm{d}}

\newcommand{\der}{\mathrm{d}}

\newcommand{\sigmap}{{ \sigma^\textrm{p}_\textrm{dip} }}

\newcommand{\dsigmap}{{\frac{\ud \sigma^\textrm{p}_\textrm{dip}}{\ud^2 \bt}}}

\newcommand{\xpom}{{x_\mathbb{P}}}

\newcommand{\ampli}{{\mathcal{N}}}
\newcommand{\A}{{\mathcal{A}}}

\title{Diffractive vector meson production in ultraperipheral
heavy ion collisions from the Color Glass Condensate}

\ShortTitle{Diffractive vector meson production from CGC}

\author{T. Lappi\\
Department of Physics, University of Jyväskylä, %
 P.O. Box 35, 40014 University of Jyv\"askyl\"a, Finland\\
 and \\
 Helsinki Institute of Physics, P.O. Box 64, 00014 University of Helsinki,
Finland \\
        E-mail: \email{tuomas.v.v.lappi@jyu.fi}}

\author{\speaker{H. Mäntysaari}\\
        Department of Physics, University of Jyväskylä, %
 P.O. Box 35, 40014 University of Jyv\"askyl\"a, Finland\\
        E-mail: \email{heikki.mantysaari@jyu.fi}}

\abstract{We compute cross sections for incoherent and coherent diffractive J/$\Psi$ and $\Psi(2S)$ production in ultraperipheral heavy ion collisions. The dipole models used in these calculations are obtained by fitting the HERA deep inelastic scattering data and compared with available electron-proton diffraction measurements. We obtain a reasonably good description of the available ALICE data. We find that the normalization of the ultraperipheral cross section has large model dependence, but the rapidity dependence is more tightly constrained.}

\FullConference{XXII. International Workshop on Deep-Inelastic Scattering and Related Subjects,\\
		28 April - 2 May 2014\\
		Warsaw, Poland}

\begin{document}

\section{Introduction}
The color glass condensate (CGC) offers a consistent framework to describe strong interactions at high energy where gluon densities become large, eventually giving a rise to nonlinear phenomena, such as gluon recombination. As the gluon density scales as $A^{1/3}$, these nonlinearities are expected to be enhanced when the target is changed from a proton to a heavy nucleus.

The CGC formalism allows one to compute many processes where the small-$x$ structure of a target is probed. These are, for example, single~\cite{Tribedy:2011aa,Albacete:2012xq,Rezaeian:2012ye,Lappi:2013zma} and double inclusive particle production~\cite{Lappi:2012nh,Albacete:2010pg,Stasto:2012ru,JalilianMarian:2012bd} in proton-proton and proton-nucleus collisions,
diffractive deep inelastic scattering (DIS)~\cite{Kowalski:2006hc,Lappi:2013am} 
and the initial state for the hydrodynamical modeling of a heavy ion 
collision~\cite{Lappi:2011ju,Schenke:2012wb,Gale:2012rq}.
The necessary ingredients for these calculations are the evolution equation for the dipole-target scattering amplitude, the BK equation~\cite{Balitsky:1995ub,Kovchegov:1999yj} (with running coupling corrections derived in Ref.~\cite{Balitsky:2006wa}), and the initial condition for the BK evolution, the dipole amplitude at initial Bjorken-$x$.

The initial condition for the evolution of the dipole amplitude can not be obtained by performing a perturbative calculations. This non-perturbative input is obtained by performing a fit to small-$x$ DIS data. The H1 and ZEUS experiments at HERA have released combined results for the proton structure functions in Refs.~\cite{Aaron:2009aa,Abramowicz:1900rp}, and the initial condition for the BK evolution has been successfully fitted to this precise data in Refs.~\cite{Lappi:2013zma,Albacete:2010sy}.

The lack of small-$x$ nuclear DIS data makes it impossible to perform a similar analysis with nuclear targets. Thus in order to obtain dipole-nucleus cross section one has to rely on modelling and use, for example, the optical Glauber model as done in Ref. \cite{Lappi:2013zma}. Future electron-ion colliders, such as LHeC~\cite{AbelleiraFernandez:2012cc} and eRHIC~\cite{Accardi:2012qut} will be able to study lepton-nucleus DIS. Before these machines are realized, it is possible to study photon-nucleus scattering is via ultraperipheral heavy ion collisions.

\section{Dipole cross section}
In the dipole picture photon-hadron scattering is described such that a (virtual) photon fluctuates to quark-antiquark pair which scatters off the hadron. The dipole-proton scattering cross section can be written as
\begin{equation}\label{eq:factbt}
\dsigmap(\bt,\rt,x) = 2T_p(\bt) \ampli(\rt,x),
\end{equation}
where $\ampli$ is the imaginary part of the forward elastic dipole-proton scattering amplitude, $\rt$ is the dipole transverse size, $\bt$ is the impact parameter and $x$ the Bjorken variable in DIS. The proton transverse thickness profile $T_p$ is assumed to be Gaussian.

The dipole amplitude $\ampli$ satisfies the BK evolution equation, and ideally one would want to use a BK evolved dipole amplitude when computing diffractive cross section. However, in these calculations an impact parameter dependent dipole amplitude is needed, and it is presently not known how one should include impact parameter dependence in the BK equation. Thus in this work we use two phenomenological parametrizations for the dipole amplitude that include a realistic impact parameter dependence.

The first dipole model considered here is the IIM model fitted to HERA $F_2$ data in Ref.~\cite{Soyez:2007kg} (for a newer fit to the combined HERA data, see Ref.~\cite{Rezaeian:2013tka}). The second parametrization used here is an eikonalized DGLAP-evolved gluon distribution (as proposed in Ref.~\cite{Bartels:2002cj}) known as the IPsat model~\cite{Kowalski:2006hc}
(see also Ref.~\cite{Rezaeian:2012ji} for a newer fit to the combined data). In the original IPsat model the impact parameter dependence is not factorized, in contrast to the IIM model. In this work we modify the IPsat model to the form of Eq.~\eqref{eq:factbt} and call it 'fIPsat'.

\section{Diffraction in DIS and ultraperipheral collisions}

Let us consider a diffractive process $\gamma^* p \to Vp$, where $V$ stands for a vector meson (e.g. a J$/\Psi$). The scattering amplitude for this process can be written as~\cite{Kowalski:2006hc}
\begin{equation}
\label{eq:diff-a}
	\A^{\gamma^* p \to Vp}(\xpom,Q^2,\Delta) = 
	\int \der^2 \rt \frac{\der z}{4\pi} [\Psi^*_V\Psi](\rt,Q^2,z) 
	 e^{-i \bt \cdot \Delta} \dsigmap(\bt,\rt,\xpom),
\end{equation}
where $\Delta$ is the momentum transfer in the process, $\xpom$ is the Bjorken $x$ in diffraction and $\Psi^*_V\Psi$ is the overlap of the virtual photon and the vector meson $V$ light cone wave functions, $z$ being the longitudinal momentum fraction carried by the quark. The vector meson wave function overlap is discussed in more detail in Sec. \ref{sec:wavef}. Physically Eq.~\eqref{eq:diff-a} means that the virtual photon splits to a quark-antiquark dipole described by the virtual photon wave function $\Psi$. This dipole then elastically scatters off the proton with cross section $ 
\sigmap$, and the scattered dipole forms a vector meson according to the vector meson wave function $\Psi_V$.
The Diffractive cross section is 
\begin{equation}
	\frac{\der \sigma^{\gamma^* p \to Vp}}{\der t} = \frac{R_g^2\left(1+\beta^2\right)}{16\pi} \left| \A^{\gamma^* p \to Vp}(\xpom,Q^2,\Delta) \right|^2,
\end{equation}
where $1+\beta^2$ accounts for the real part of the scattering amplitude and $R_g^2$ corrects for the skeweness effect~\cite{Shuvaev:1999ce,Martin:1999wb} calculated as in Ref.~\cite{Watt:2007nr}.

To extend the dipole cross section from protons to nuclei we take the independent scattering approximation and write the $S$-matrix as
\begin{equation}
	S_A(\rt,\bt,\xpom) = \prod_{i=1}^A S_p(\rt, \bt - \bt_i,\xpom).
\end{equation}
When considering diffraction off a heavy nucleus, there are two separate event types. First, in \emph{coherent diffraction} the target nucleus remains intact, and the cross section is obtained by averaging the amplitude over the nucleon configurations from the Woods-Saxon distribution: $\der \sigma \sim \left|\langle \A \rangle_N \right|^2$. It is also possible that the target nucleus breaks up, but the event remains diffractive. This is called \emph{incoherent diffraction} and the cross section is calculated as a variance $\der \sigma \sim \left \langle |\A|^2\right\rangle_N - \left|\langle \A\rangle_N\right|^2$. For details on the calculations of the diffractive cross sections, see Refs.~\cite{Kowalski:2003hm,Lappi:2010dd}.

In ultraperipheral heavy ion collisions, when the impact parameter is larger than twice the nuclear radius, the nuclei do not touch each other and the strong interactions are suppressed. The dominant interaction channel is electromagnetic, and the events can be described such that one of the colliding nuclei acts as source of (virtual) photons that scatter off the second nucleus. The cross section for a production of a vector meson $V$ with rapidity $y$ can be written as
\begin{equation}
	\frac{\der \sigma^{A_1A_2 \to V A_1A_2}}{\der y} = n^{A_2}(y) \sigma^{\gamma A_1}(y) + n^{A_1}(-y)\sigma^{\gamma A_2}(-y),
\end{equation}
where $n^{A_i}$ is the photon flux generated by nucleus $i$. For more details and the expression for the photon flux, see Ref.~\cite{Bertulani:2005ru}.

\section{Vector meson wave functions}
\label{sec:wavef}
The virtual photon wave function ($\gamma^* \to q\bar q$ splitting) can be computed from QED. On the other hand the formation of a vector meson from a quark-antiquar dipole requires some modelling.  There exists a few different parametrizations for the virtual photon-vector meson wave function overlap. In this work we calculate the J/$\Psi$ production  using the so called Boosted Gaussian and Gaus-LC wave functions from Ref.~\cite{Kowalski:2006hc}.

We shall also consider production of $\Psi(2S)$, which is an excited state of J/$\Psi$. In order to obtain the wave function overlap for the $\Psi(2S)$ we follow the procedure used in Ref.~\cite{Cox:2009ag} for $\Upsilon$: the wave function is required to be orthogonal to that of the J/$\Psi$, the decay width to electrons and correct normalization are required. Following the notation of Ref.~\cite{Cox:2009ag}, we obtain the parameters $R_{2s}=1.851$ GeV$^{-1}$, $\alpha_{2s,1}=-0.55816$ and $N_{2}=0.7394$ for the $\Psi(2S)$ Boosted Gaussian wave function with transverse polarization.

\section{Results}

\begin{figure}
\begin{minipage}[t]{0.48\linewidth}
\centering
\includegraphics[width=1.05\textwidth]{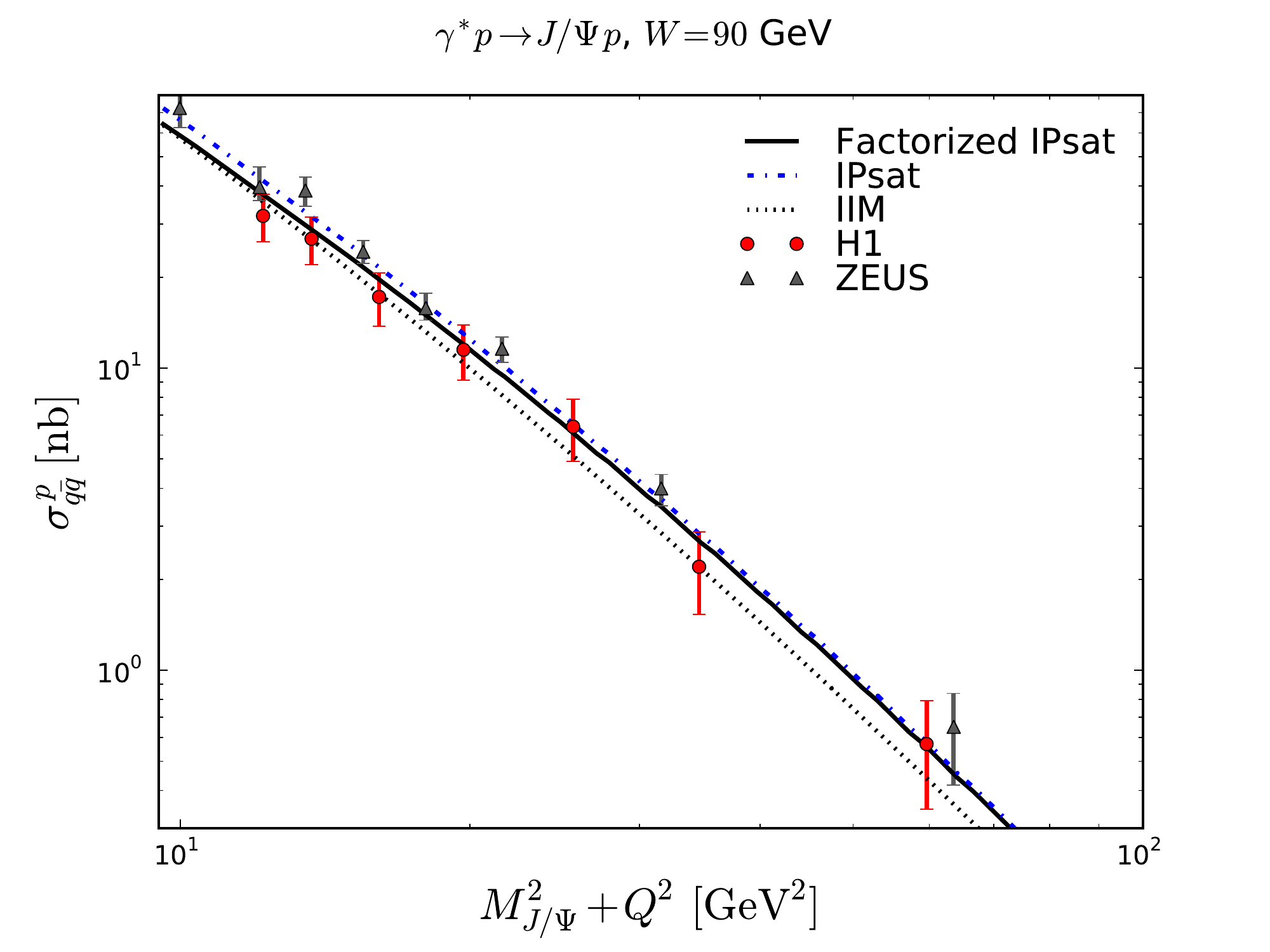} 
\caption{Total diffractive J/$\Psi$ production cross section computed using the Boosted Gaussian wave function compared to HERA data~\cite{Chekanov:2004mw,Aktas:2005xu}.}
\label{fig:totxs} 
\end{minipage}
\hspace{0.5cm}
\begin{minipage}[t]{0.48\linewidth}
\centering
\includegraphics[width=1.05\textwidth]{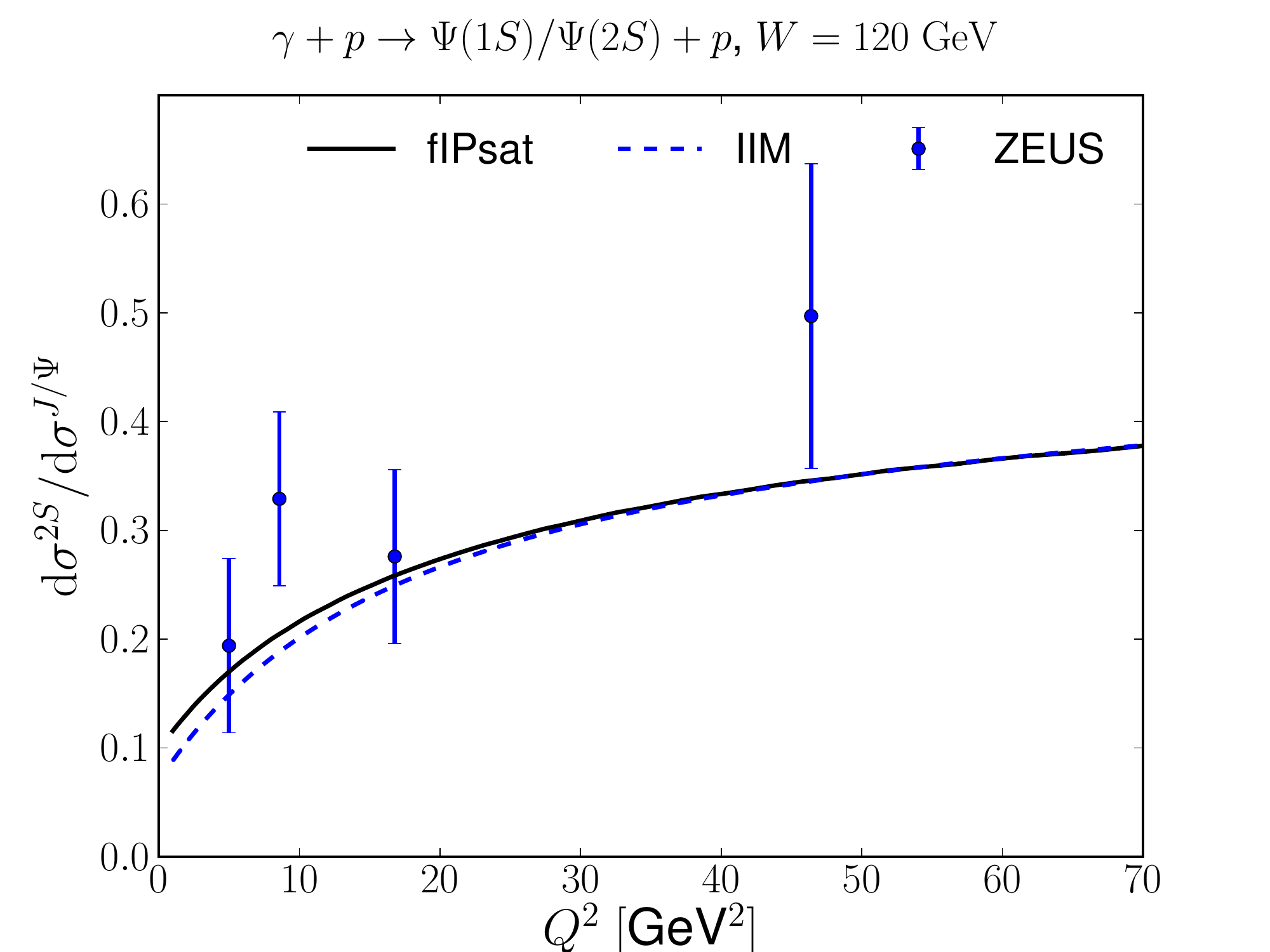} 
\caption{Diffractive $\Psi(2S)$ and J/$\Psi$ production ratio in $\gamma^*p$ collisions as a function of $Q^2$ compared to preliminary ZEUS data~\cite{kovalchuk2s1sratiodis}.}
\label{fig:ratio_qsqr} 
\end{minipage}
\end{figure}

\begin{figure}
\begin{minipage}[t]{0.48\linewidth}
\centering
\includegraphics[width=1.05\textwidth]{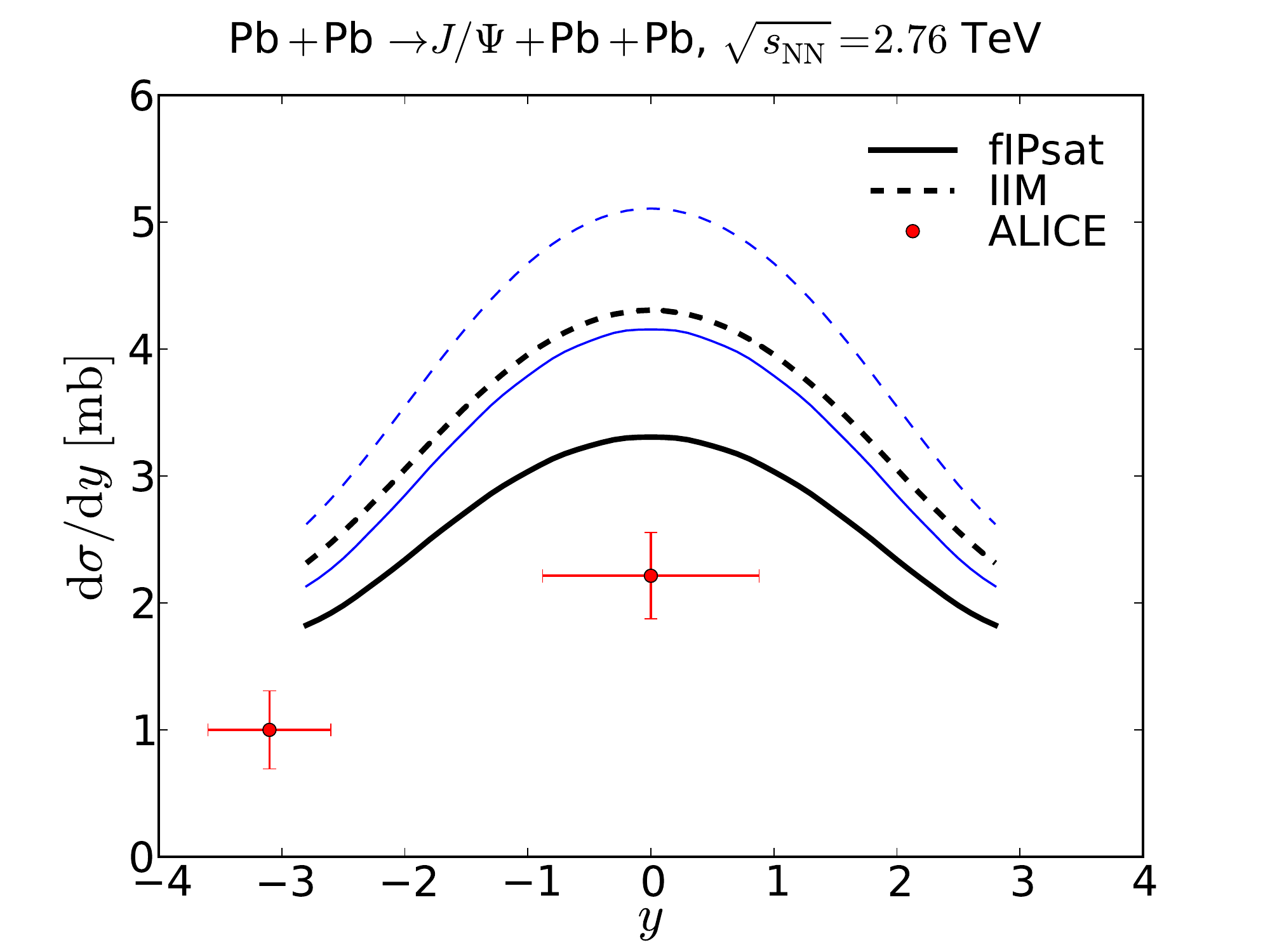} 
\caption{Coherent diffractive J/$\Psi$ production in ultraperipheral lead-lead collisions computed using the fIPsat and IIM dipole cross sections and Boosted Gaussian (thin lines) and Gaus-LC (thick lines) wave functions. ALICE data from Ref.~\cite{Abbas:2013oua}.}
\label{fig:cohaa} 
\end{minipage}
\hspace{0.5cm}
\begin{minipage}[t]{0.48\linewidth}
\centering
\includegraphics[width=1.05\textwidth]{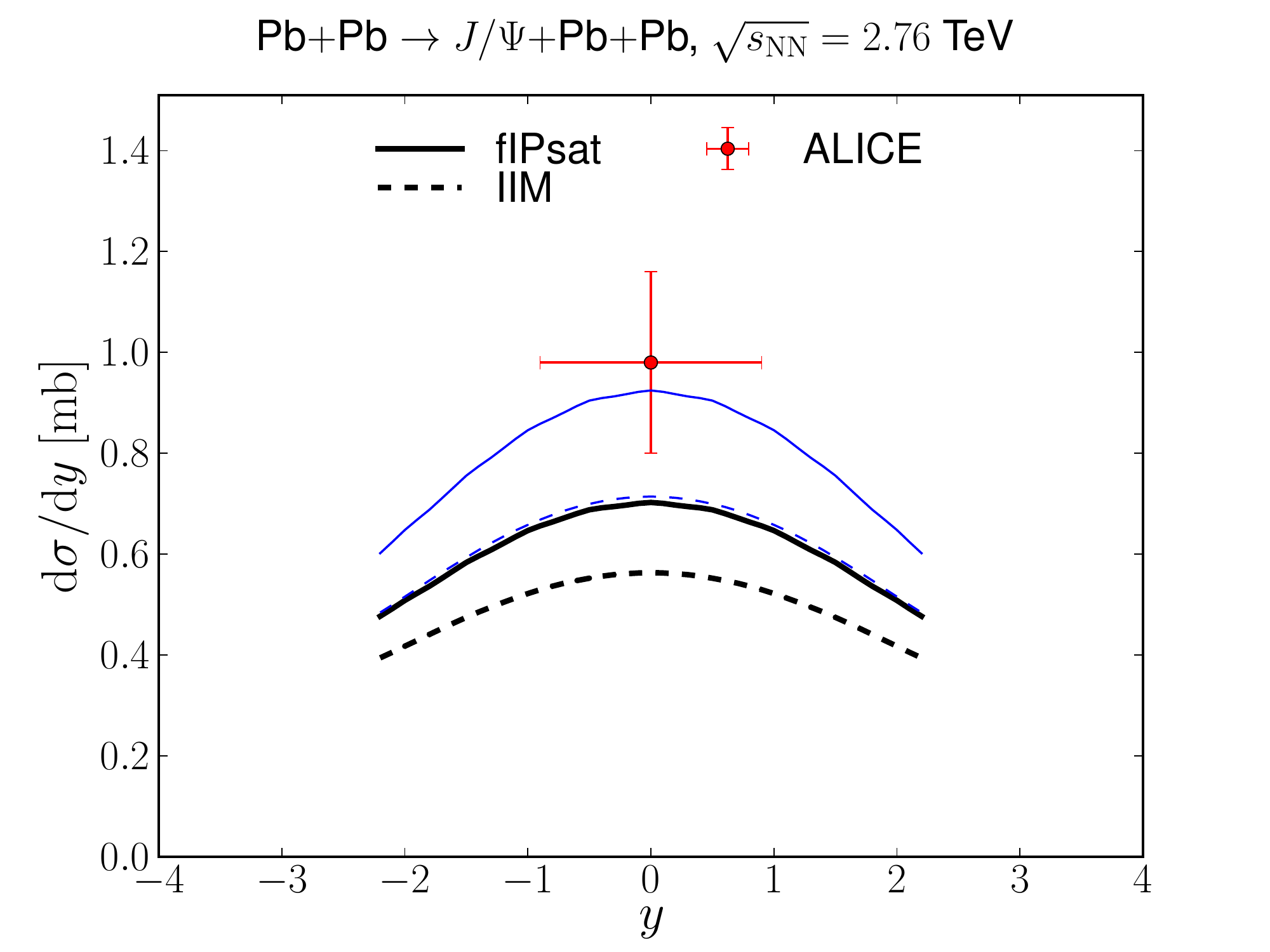} 
\caption{Incoherent diffractive J/$\Psi$ production in ultraperipheral lead-lead collisions computed using the fIPsat and IIM dipole cross sections and Boosted Gaussian (thin lines) and Gaus-LC (thick lines) wave functions. ALICE data from Ref.~\cite{Abbas:2013oua}.}
\label{fig:incohaa} 
\end{minipage}
\end{figure}

In order to validate our model we compute diffractive J/$\Psi$ production cross section in $\gamma^*p$ collisions and compare with the HERA data \cite{Chekanov:2004mw,Aktas:2005xu}. The results obtained with original IPsat parametrization, factorized version of it and the IIM dipole cross sections are shown in Fig.~\ref{fig:totxs}. We also compute the ratio of $\Psi(2S)$ and J/$\Psi$ production diffractive DIS and compare with the preliminary ZEUS data shown in this Conference~\cite{kovalchuk2s1sratiodis}. The result is shown in Fig. \ref{fig:ratio_qsqr}. The agreement with the electron-proton data is relatively good, taking into account the experimental uncertainties.

Cross sections for coherent and incoherent diffractive J/$\Psi$ production cross sections are shown in Figs.~\ref{fig:cohaa} and \ref{fig:incohaa} compared to the ALICE data~\cite{Abbas:2013oua}. The absolute normalization depends quite strongly on the dipole cross section and wave function model, but the rapidity distribution is a more solid prediction. We emphasize that all the parameters in the calculation come from fits to HERA DIS data, combined with the Woods-Saxon distribution.

The ALICE collaboration has also measured coherent $\Psi(2S)$ production in ultraperipheral heavy ion collisions at midrapidity at $\sqrt{s}=2.76$ TeV, obtaining $\der \sigma/\der y = 0.83 \pm 0.19$ mb~\cite{nystrand2s}. Using fIPsat or IIM dipole cross sections with the Boosted Gaussian wave function we get $\der \sigma/\der y = 0.64 \dots 0.65$ mb. 

As a conclusion we note that it is possible to consistently describe coherent and incoherent diffractive vector meson production in ultraperipheral heavy ion collisions. The theoretical uncertainties on the absolute normalization are currently relatively large, but a simultaneous description of both diffractive event classes can help to constrain theoretical models.

\section*{Acknowledgements}
We thank J. Nystrand, J. Tomaszewska and M.V.T. Machado for discussions.
This work has been supported by the Academy of Finland, projects 133005, 
267321, 273464. H.M. is supported by the Graduate School of 
Particle and Nuclear Physics.

\bibliographystyle{h-physrev4mod2}
\bibliography{../../refs}

\end{document}